%% file: article.tex
\documentclass[12pt]{article}
\usepackage{epsfig}
\usepackage{float}
\usepackage{subcaption}
\usepackage{amsfonts} 
\usepackage{graphicx} 
\usepackage{amsmath}
\usepackage{chngcntr}
\usepackage{wrapfig}
\usepackage{caption}

\input econfmacros.tex
\def\Title#1{\begin{center} {\Large {\bf #1} } \end{center}}

\begin{document}

\Title{Search for Massive Bosons Decaying to W$\gamma$ and Z$\gamma$ Using the ATLAS Detector}

\bigskip\bigskip


\begin{raggedright}  

{\it Wei Tang\\
Department of Physics\\
Duke University}
\bigskip\bigskip
\end{raggedright}

\tableofcontents
\clearpage

\section{Introduction}
\begin{figure}[H]
\begin{center}
\epsfig{file=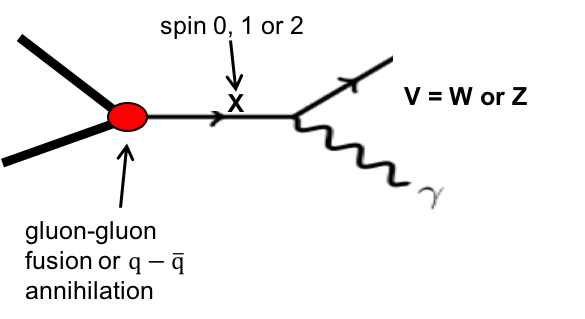, scale=0.8}
\caption[Caption for LOF]{X$\rightarrow$V$\gamma$}
\label{decay-channels}
\end{center}
\end{figure}

The Standard Model is by far the most encompassing physics theory. With the recent discovery of the Higgs Boson, the Standard Model has performed extremely well against experimental data. However, the theory is intrinsically not complete, it does not address physical phenomenon such as gravity and dark matter. Many proposals for physics beyond the Standard Model predict new massive bosons from additional gauge fields or extended Higgs sectors.\\
At CERN, we analyze proton-proton collision experimental data to search for evidence of new particles. This paper examines possible decay channels of an unknown massive boson X decaying into V plus a gamma, where V can be either a W or a Z boson. The production mechanism can be either gluon-gluon fusion or $q-\bar{q}$ annihilation. Figure \ref{decay-channels} illustrates the possible production and decay channels.\\
The data being analyzed in this paper is collected by the ATLAS detector at the CERN Large Hadron Collider. The data comes from proton-proton collisions with a center of mass energy of 13TeV, and has a total integrated luminosity of 36.1 $fb^{-1}$. The data is collected using a trigger based upon a photon with transverse energy greater than 140 GeV.\\
This paper describes preliminary results from a generic search for new massive X bosons. The ultimate goal of our search is to explore the X mass range from 250 GeV to highest energy attained in 13TeV p-p collision.

\section{Leptonic Decay}
The boson decay products from the unknown massive boson X can further undergo either leptonic or hadronic decay. In the leptonic decay, the Z boson further decays into a pair of either electron and positron or $\mu^+$ $\mu^-$.
\subsection{Event Selections}
The leptonic decay events were selected using single electron and muon triggers with a nominal transverse momentum larger than 26 GeV, supplemented by di-lepton triggers with lower thresholds.\\
The Z bosons which decay to a pair of leptons are selected using well measured, isolated electron and muon pairs with invariant mass within $\pm$15 GeV of the invariant mass of Z bosons.\\
The photons are required to be isolated with pseudo rapidity $|\eta|<$1.37 or 1.52$<|\eta|<$2.37 and transverse momentum $P_T>$ 10 GeV. These requirements ensure that we are using the detector region with highest granularity.\\
The final Z $\gamma$ event selection requires the photon to have transverse momentum greater than 30\% of the combined Z $\gamma$ invariant mass.
\subsection{Signal Simulation}
\begin{figure}[H]
\begin{center}
\epsfig{file=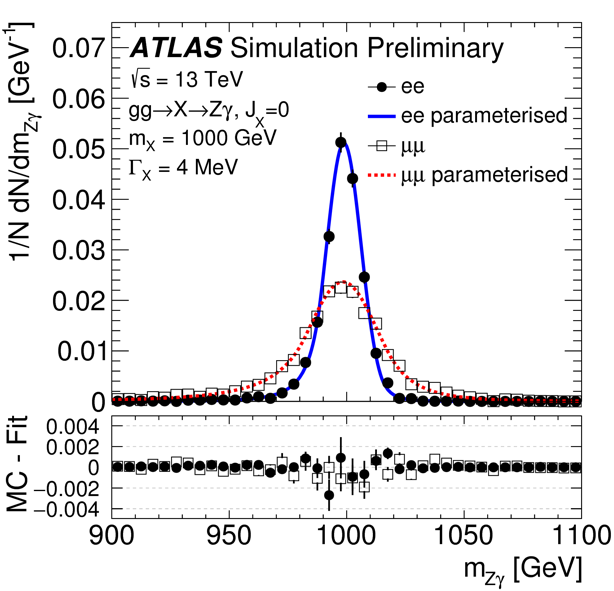, scale=0.6}
\caption[Caption for LOF]{Z$\gamma$ Invariant Mass Simulation}
\label{signal-simulation}
\end{center}
\end{figure}

\begin{figure}[H]
\begin{center}
\epsfig{file=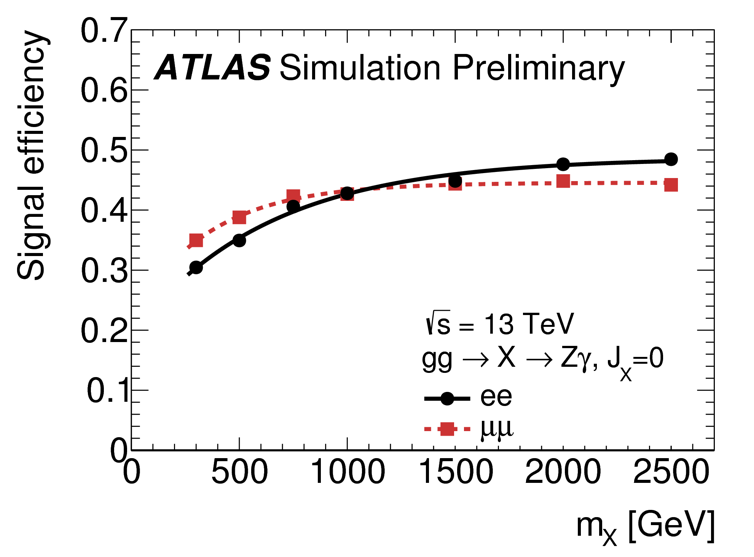, scale=0.6}
\caption[Caption for LOF]{Simulation Signal Efficiency}
\label{signal-efficiency}
\end{center}
\end{figure}

In the signal simulation, unknown massive boson X decays with a narrow width to a Z boson and $\gamma$. The narrow width chosen is 4MeV. Figure \ref{signal-simulation} on page \pageref{signal-simulation} shows the Z$\gamma$ combined invariant mass distribution with all selection cuts applied.\\
Figure \ref{signal-efficiency} on page \pageref{signal-efficiency} shows the total signal efficiency as a function of $M_X$. The signal efficiency of all the selection cuts combined range from 30-50\% for increasing $M_X$.\\
Event samples are generated using POWHEG-BOX interfaced with Pythia, with $M_X$ ranges from 200 GeV to 2.4 TeV. 
\subsection{Background Simulation}
There are two sources of background. The dominant background is the Standard Model production of a Z boson plus a photon, with smaller contributions from Z + jets with the hadronic jet misidentified as a photon.\\
\begin{figure}
\centering
\begin{subfigure}{.5\textwidth}
  \centering
  \includegraphics[width=.8\linewidth]{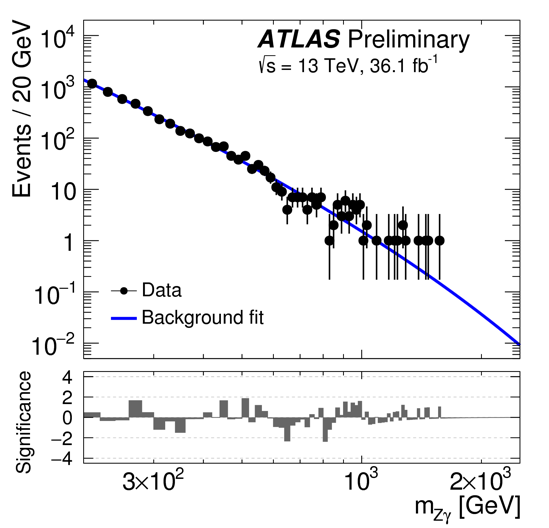}
  \caption{$M_{Z\gamma}$ Background Only Fit}
  \label{background}
\end{subfigure}%
\begin{subfigure}{.5\textwidth}
  \centering
  \includegraphics[width=1\linewidth]{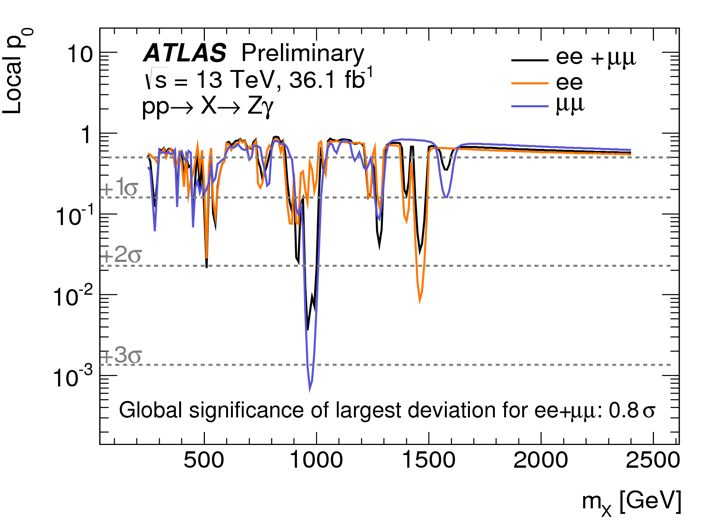}
  \caption{p-value of the $M_{Z\gamma}$ Distribution}
  \label{p-value}
\end{subfigure}
\caption{Significance of $M_{Z\gamma}$ Observations}
\end{figure}
The Z$\gamma$ invariant mass data points are plotted with the background-only fit, as shown in Figure \ref{background} on page \pageref{background}. The solid blue line is the background prediction and black dots are the actual data. The background Z$\gamma$ mass spectrum is smoothly falling and can be parameterized with $f(x)\approx(1-x^{1/3})^{p_1}x^{p_2}$ with $x = M_{Z_\gamma} / \sqrt{s}$. Figure \ref{p-value} on page \pageref{p-value} shows the local p-value of the $M_{Z\gamma}$ data with respect to the background-only hypothesis. By comparing the data and the background, the largest local deviation is around $M_x=268GeV$ where the local significance is 2.2 $\sigma$.\\
\begin{figure}[H]
\begin{center}
\epsfig{file=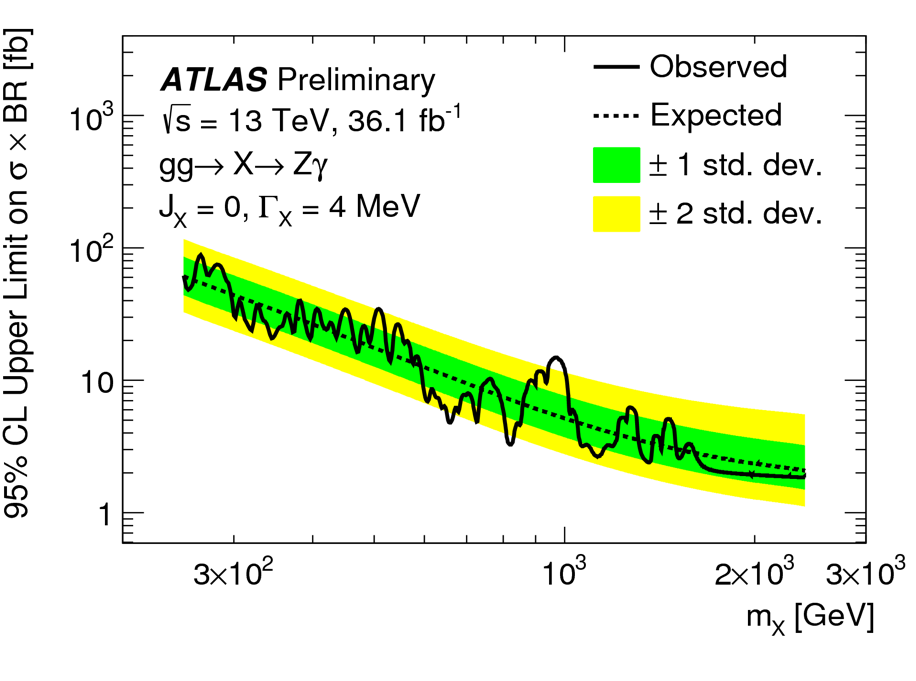, scale=0.6}
\caption[Caption for LOF]{Confidence Level Limit}
\label{CL_limit}
\end{center}
\end{figure}
Upper limits on $\sigma(pp\rightarrow X)\times BR(X\rightarrow Z\gamma)$ are set using a profile likelihood method. Figure \ref{CL_limit} on page \pageref{CL_limit} shows the observed (solid lines) and median expected (dashed lines) 95\% confidence level limits on the product of the production cross section times the branching ratio for the decay to a Z boson and a photon of a narrow scalar boson X,  $\sigma(pp\rightarrow X)\times BR(X\rightarrow Z\gamma)$. The green and yellow band indicate 1 and 2 standard deviation intervals. A modified frequentist (CLs) method is used to set upper limits on the product, by identifying the value of $\sigma\times$BR for which CLs=0.05.\\
The observed $\sigma\times$BR limits vary from 88 fb at $M_X$ = 250  GeV to 2.8 fb at $M_X$ = 2.40 TeV. There is not a significant deviation from the observed $\sigma\times$BR limits.

\section{Hadronic Decay}
\begin{figure}[H]
\begin{center}
\epsfig{file=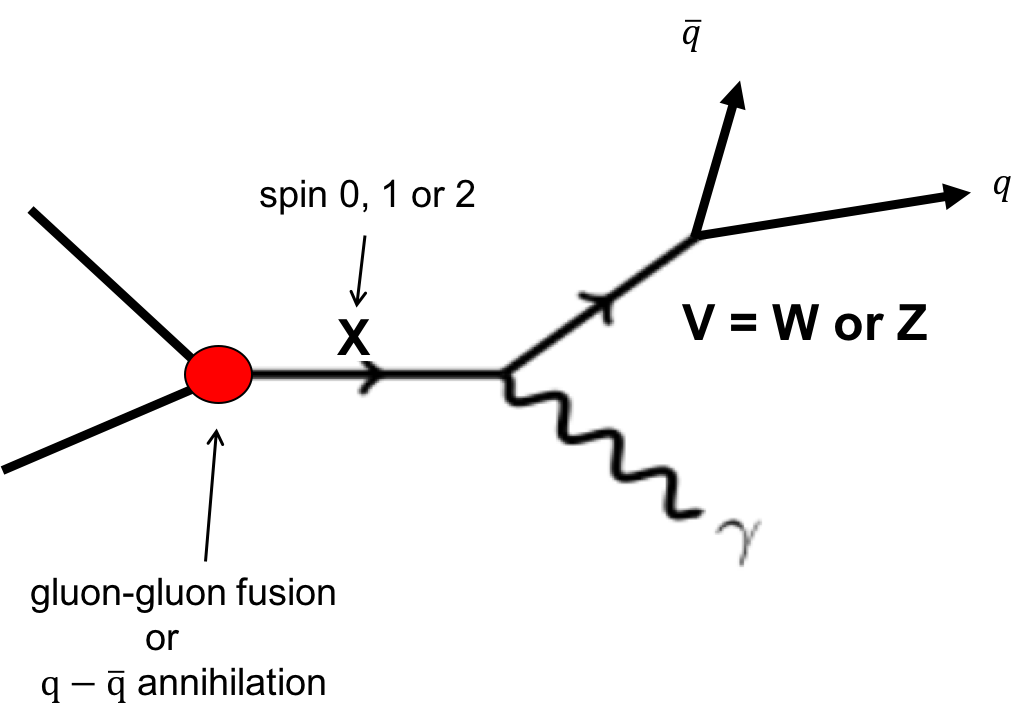, scale=0.45}
\caption[Caption for LOF]{Hadronic Decay Channel}
\label{hadronic}
\end{center}
\end{figure}

The hadronic decays of the W/Z bosons have much higher branching ratio than the leptonic decays. The search sensitivity for $X\rightarrow W/Z + \gamma$ can be improved by capturing a higher fraction of the W/Z bosons from their hadronic decays. The branching ratio of $W/Z\rightarrow q\bar{q}$ is around 70\%. Figure \ref{hadronic} on page \pageref{hadronic} shows the production and decay channels.
\subsection{Event Selections}
\begin{figure}[H]
\begin{center}
\epsfig{file=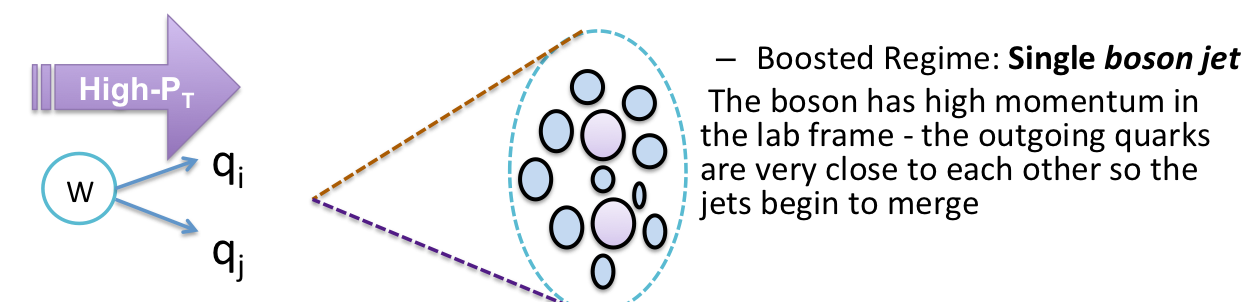, scale=0.6}
\caption[Caption for LOF]{Identification of W/Z Boosted Jets}
\label{boosted_jet}
\end{center}
\end{figure}

This search uses a selection of events based upon a photon trigger with transverse momentum $P_T>140GeV$, followed by identification of highly boosted $W/Z(q\bar{q})$ bosons using fat jets with cone size R = 1.0. A jet substructure variable $D_2^{(\beta=1)}$ is used to identify fat jets with 2-jet substructure, to select hadronically decaying W/Z bosons while suppressing jets from single quarks or gluons. Figure \ref{boosted_jet} on page \pageref{boosted_jet} shows the $q\bar{q}$ decays from a single W/Z boson. Due to the high momentum of the W/Z boson, the $q\bar{q}$ decay products are very close together and their subsequent shower merge.\\
The spin 0 gluon-gluon fusion production with $Z\gamma$ decay signals are generated using the POWHEG-BOX generator. Both spin 1 $X\rightarrow W(q\bar{q}) + \gamma$ and spin 2 $X\rightarrow Z(q\bar{q}) + \gamma$ signals are generated using the MadGraph generator.\\
The selection efficiency of the fat-jet identification of the W and Z bosons depends on the $\Delta R(q\bar{q})$ decay distribution, and is therefore sensitive to the polarization of the W/Z bosons.    This paper studies the signal efficiency as a function of the X boson mass using preliminary kinematic selection criteria. The kinematic selection criteria are:
\begin{enumerate}
\item $|\eta_{(q\bar q)}|<2$
\item $PT_{(q\bar{q})}>200GeV$
\item $PT_\gamma>250GeV$
\item $|\eta_\gamma|<1.37$
\end{enumerate}
\subsection{Selection Efficiencies}
\begin{figure}
\centering
\begin{subfigure}{.5\textwidth}
  \centering
  \includegraphics[width=1\linewidth]{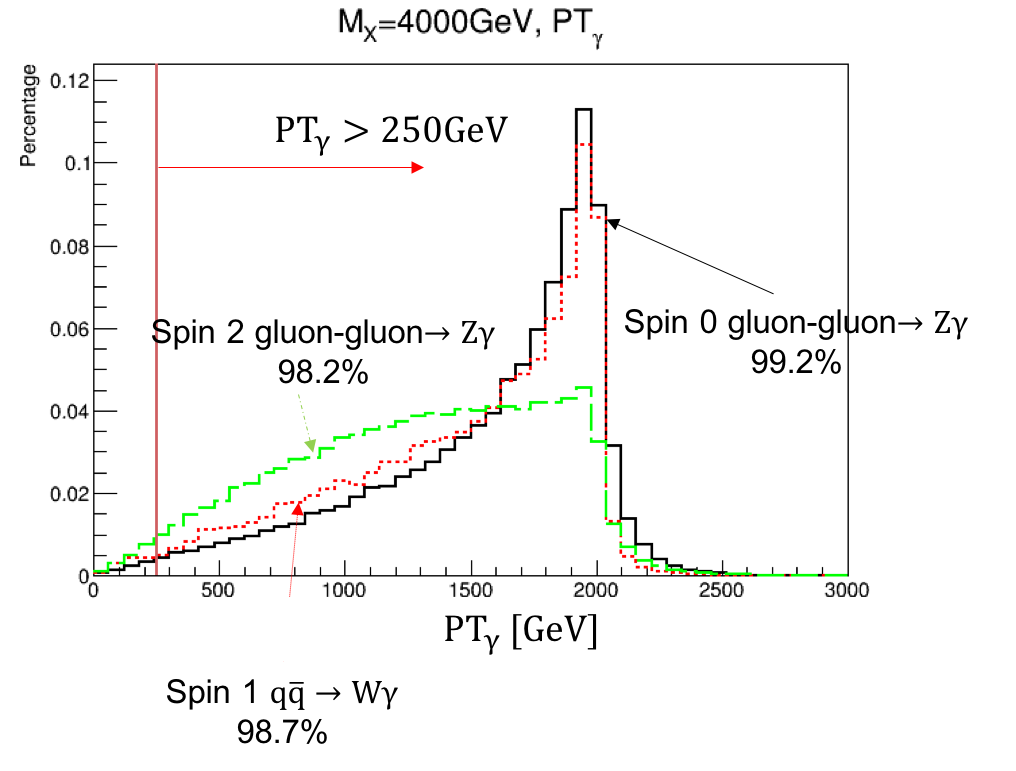}
  \caption{$PT_\gamma>250GeV$ Selection}
  \label{pt_gamma_cut}
\end{subfigure}%
\begin{subfigure}{.5\textwidth}
  \centering
  \includegraphics[width=1\linewidth]{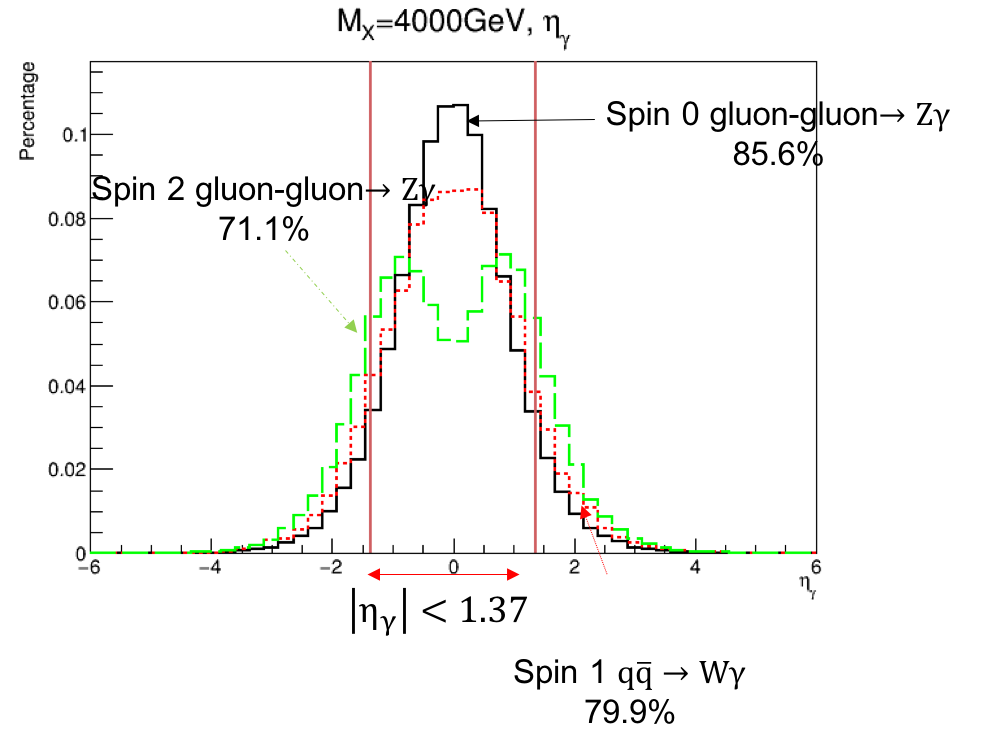}
  \caption{$|\eta_\gamma|<1.37$ Selection}
  \label{eta_gamma_cut}
\end{subfigure}
\caption{$\gamma$ Kinematics Selection Cuts}
\label{gamma_cuts}
\end{figure}

\begin{figure}[H]
\begin{center}
\epsfig{file=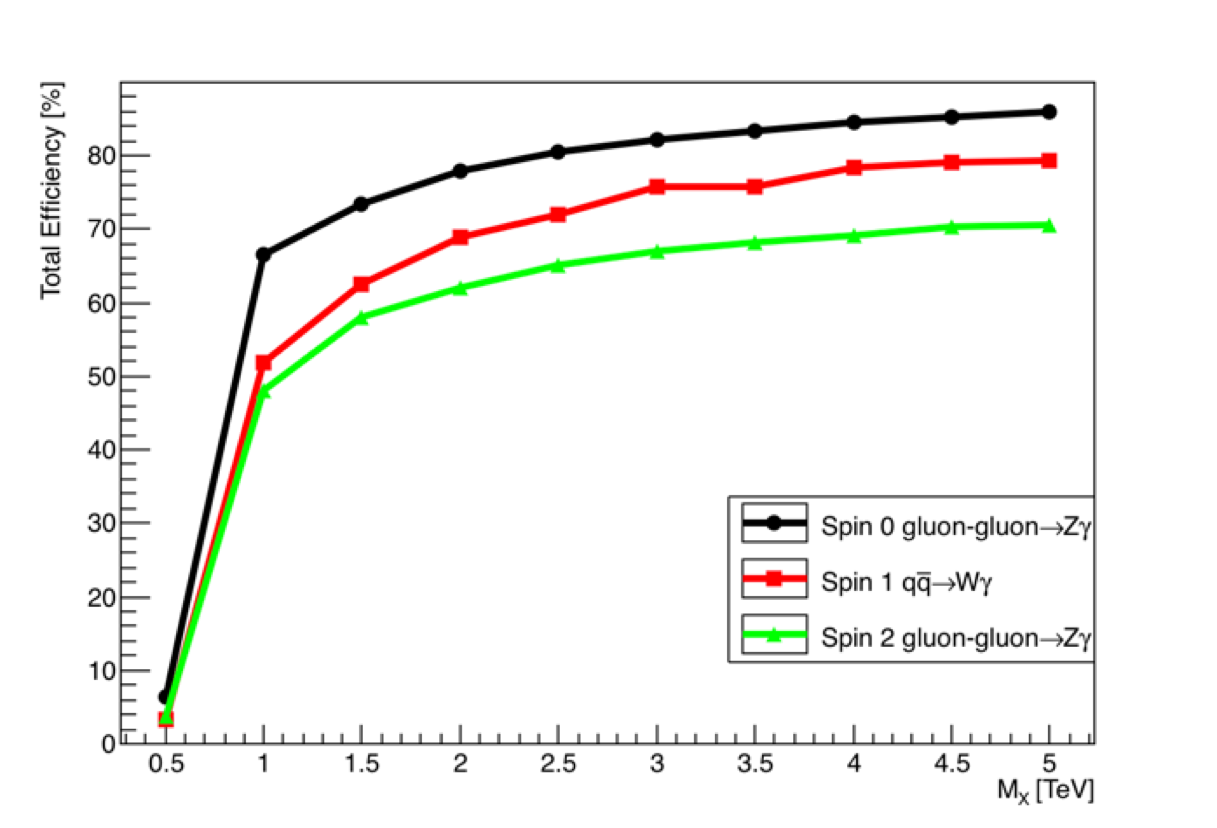, scale=0.45}
\caption[Caption for LOF]{Signal Efficiency from Combined Kinematic Selection Cuts on All Three Production Channels}
\label{total_efficiency}
\end{center}
\end{figure}

Figures \ref{gamma_cuts} and \ref{qq_cuts} show the selection efficiencies for the preliminary kinematic cuts applied separately on the signal events. The examples shown here are for $M_X$=4TeV. The same selection cuts are applied to signal events of all $M_X$ slices. Figure \ref{total_efficiency} shows the total signal efficiencies from applying the combined kinematic selection cuts as a function of $M_X$ for the three different hadronic channels. The signal efficiencies increase abruptly from low $M_X$ to high masses.

\begin{figure}
\centering
\begin{subfigure}{.5\textwidth}
  \centering
  \includegraphics[width=1\linewidth]{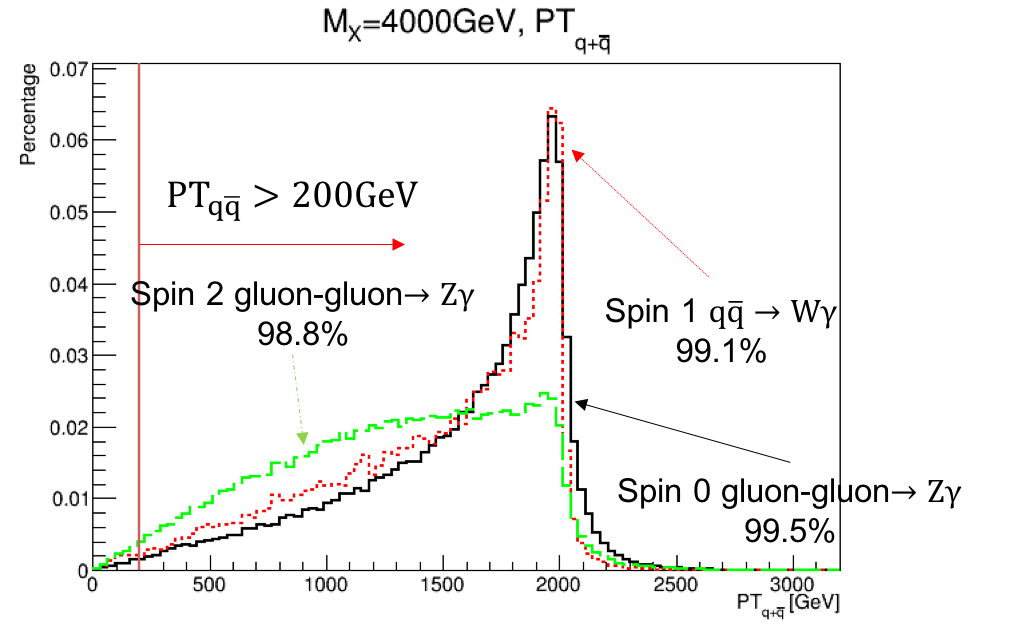}
  \caption{$PT_\gamma>250GeV$ Selection}
  \label{pt_gamma_cut}
\end{subfigure}%
\begin{subfigure}{.5\textwidth}
  \centering
  \includegraphics[width=1\linewidth]{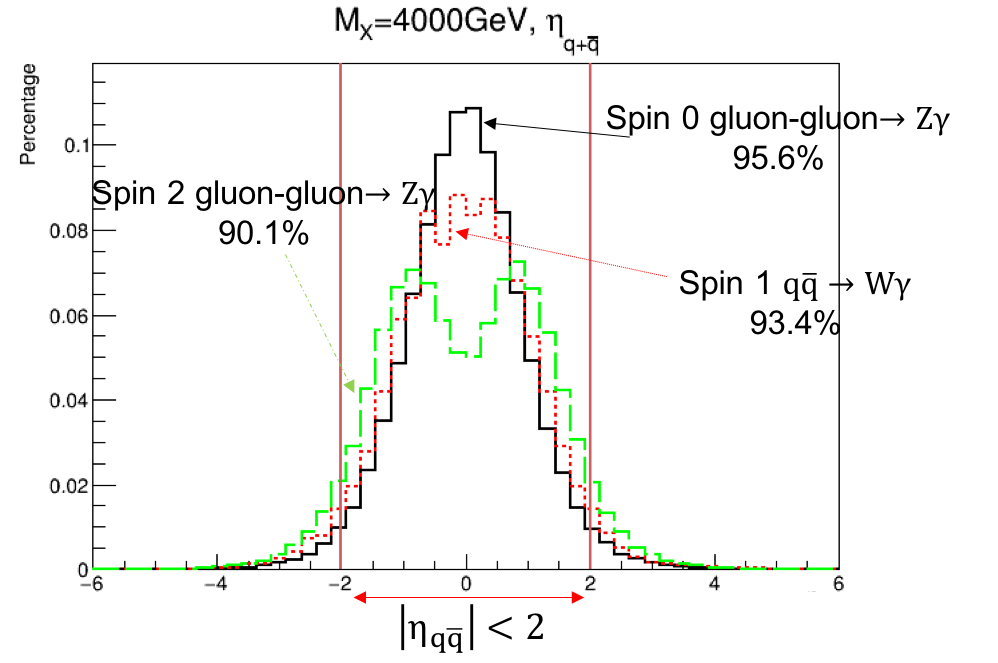}
  \caption{$|\eta_\gamma|<1.37$ Selection}
  \label{eta_gamma_cut}
\end{subfigure}
\caption{$q\bar{q}$ Kinematics Selection Cuts}
\label{qq_cuts}
\end{figure}

\subsection{Decay Polarizations}
\begin{figure}[H]
\begin{center}
\epsfig{file=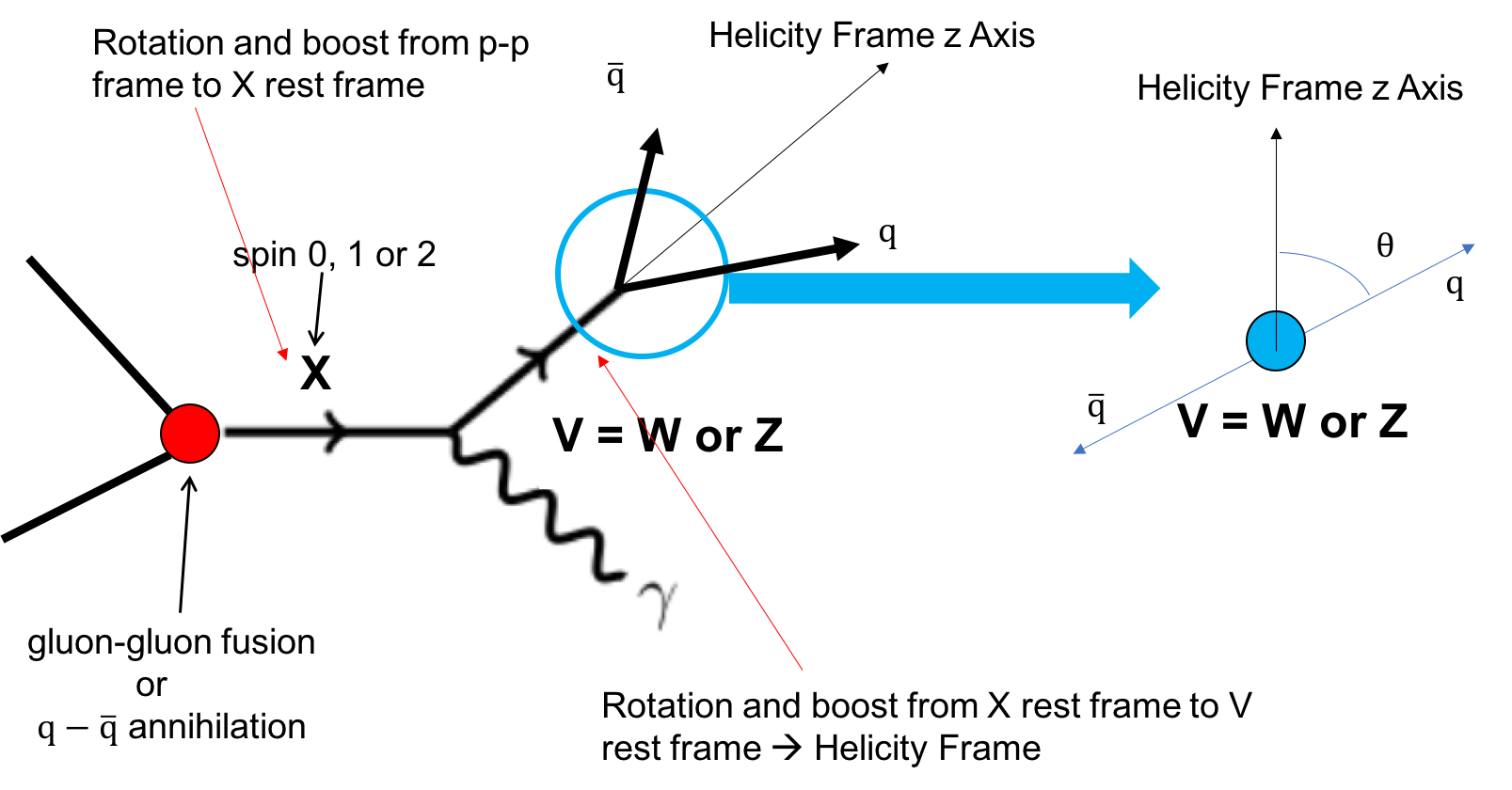, scale=0.4}
\caption[Caption for LOF]{Helicity Frame Definition}
\label{helicity}
\end{center}
\end{figure}

\begin{figure}[H]
\begin{center}
\epsfig{file=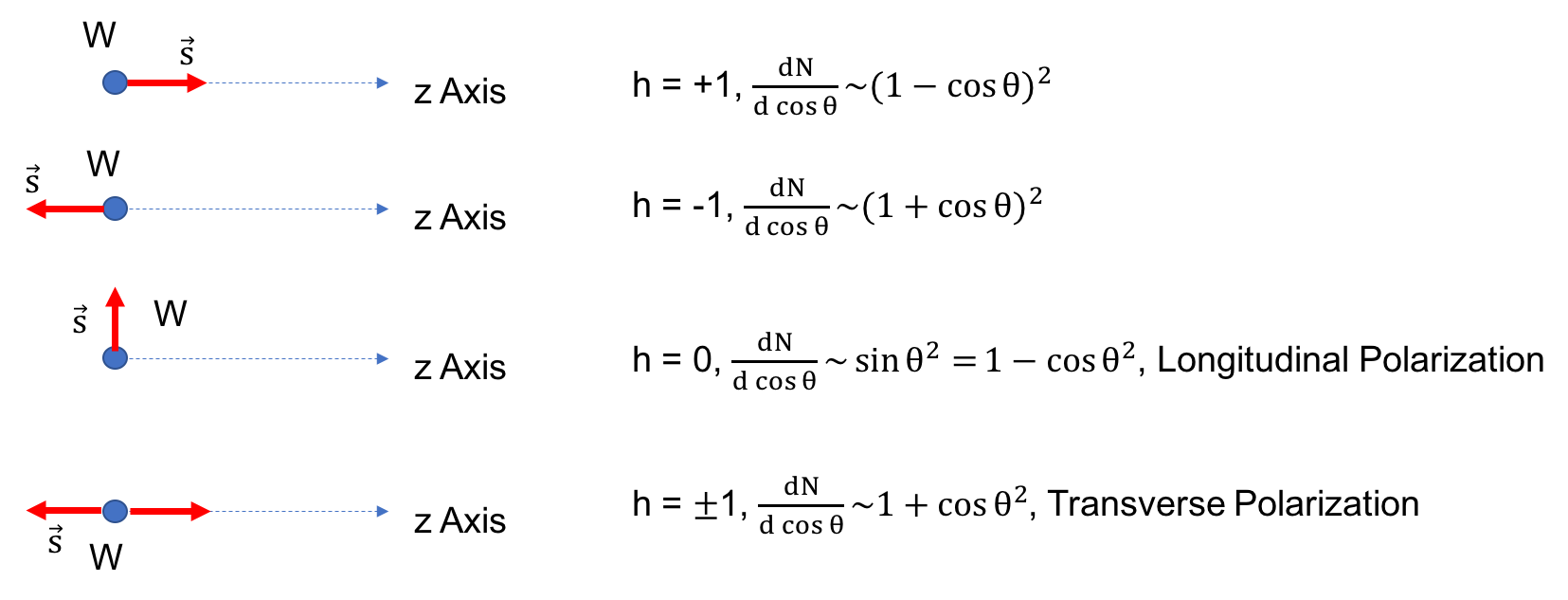, scale=0.5}
\caption[Caption for LOF]{Functional forms of decay polarizations for different W/Z production channels}
\label{polarization_interpretation}
\end{center}
\end{figure}

\begin{figure}
\centering
\begin{subfigure}{.4\textwidth}
  \centering
  \includegraphics[width=1\linewidth]{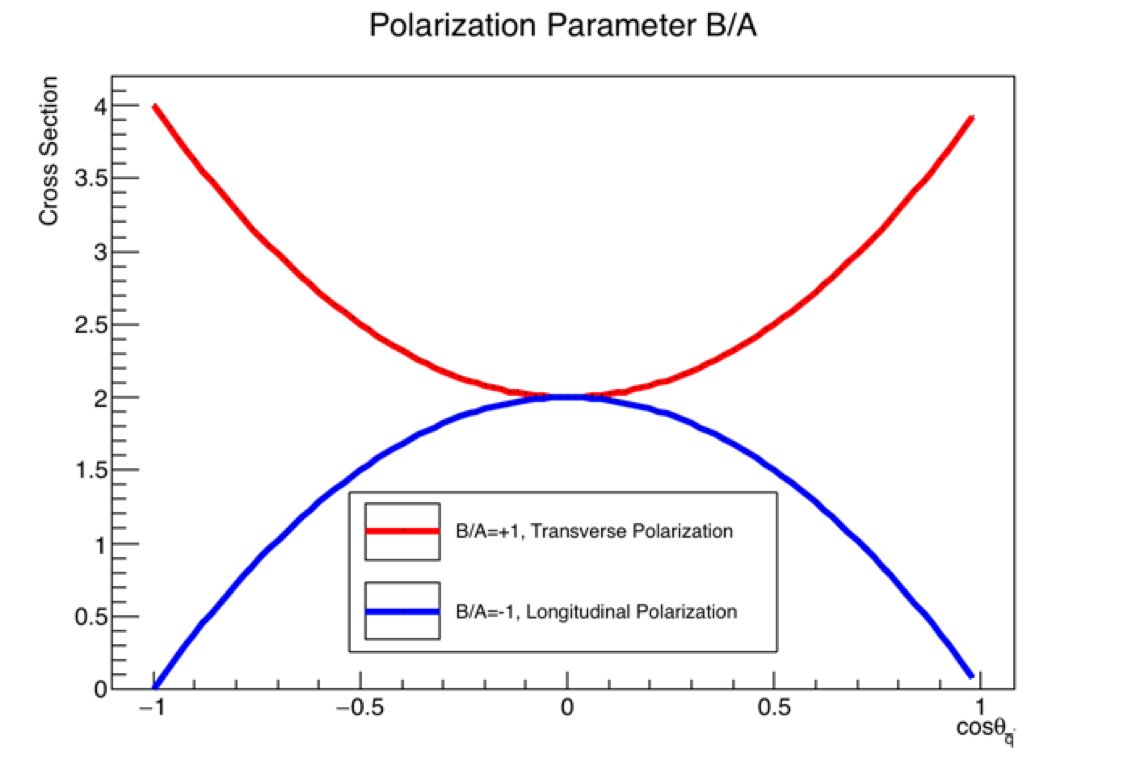}
  \caption{$\cos\theta$ Distributions for purely longitudinal (blue) and purely transverse (red) polarizations}
  \label{polarization_parameter}
\end{subfigure}%
\begin{subfigure}{.6\textwidth}
  \centering
  \includegraphics[width=1.2\linewidth]{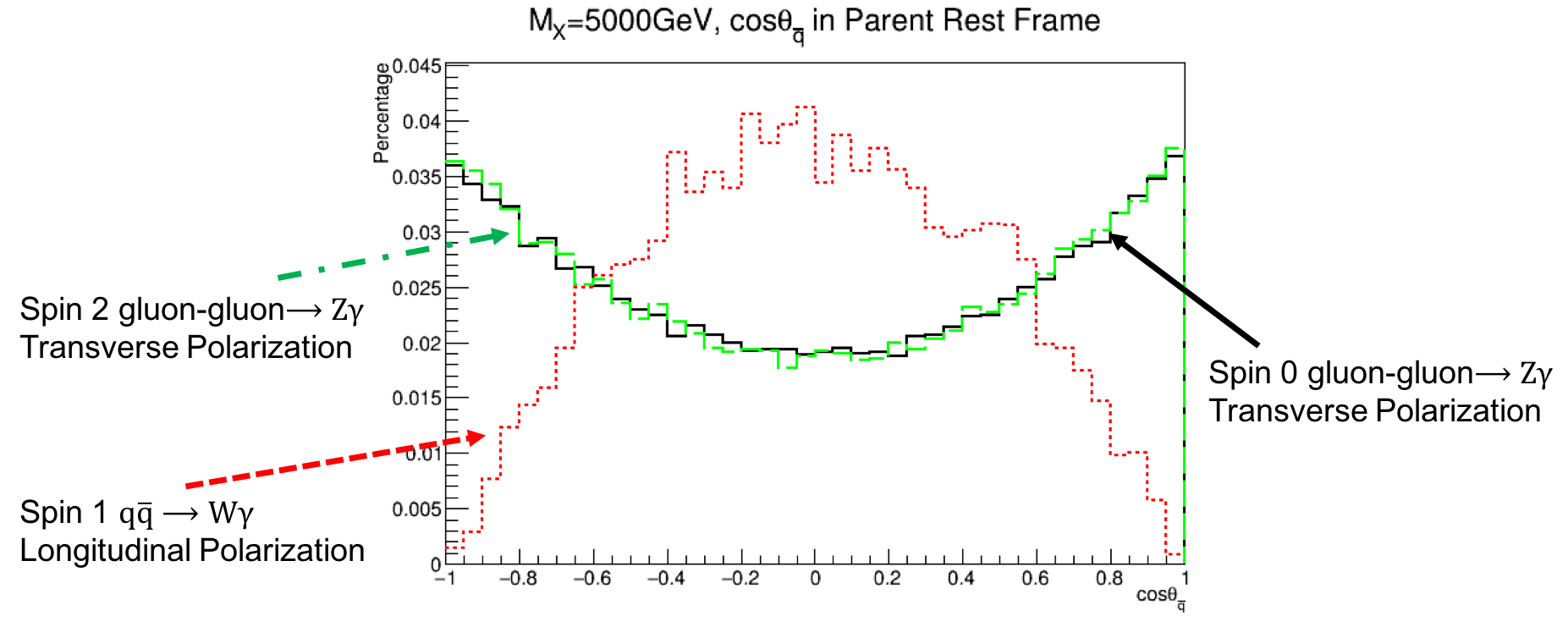}
  \caption{$\cos\theta$ Distributions for three channels at $M_X$=5TeV}
  \label{polarization_example}
\end{subfigure}
\caption{Theoretical $\cos\theta$ distributions and an example distribution}
\label{qq_cuts}
\end{figure}

\begin{figure}[H]
\begin{center}
\epsfig{file=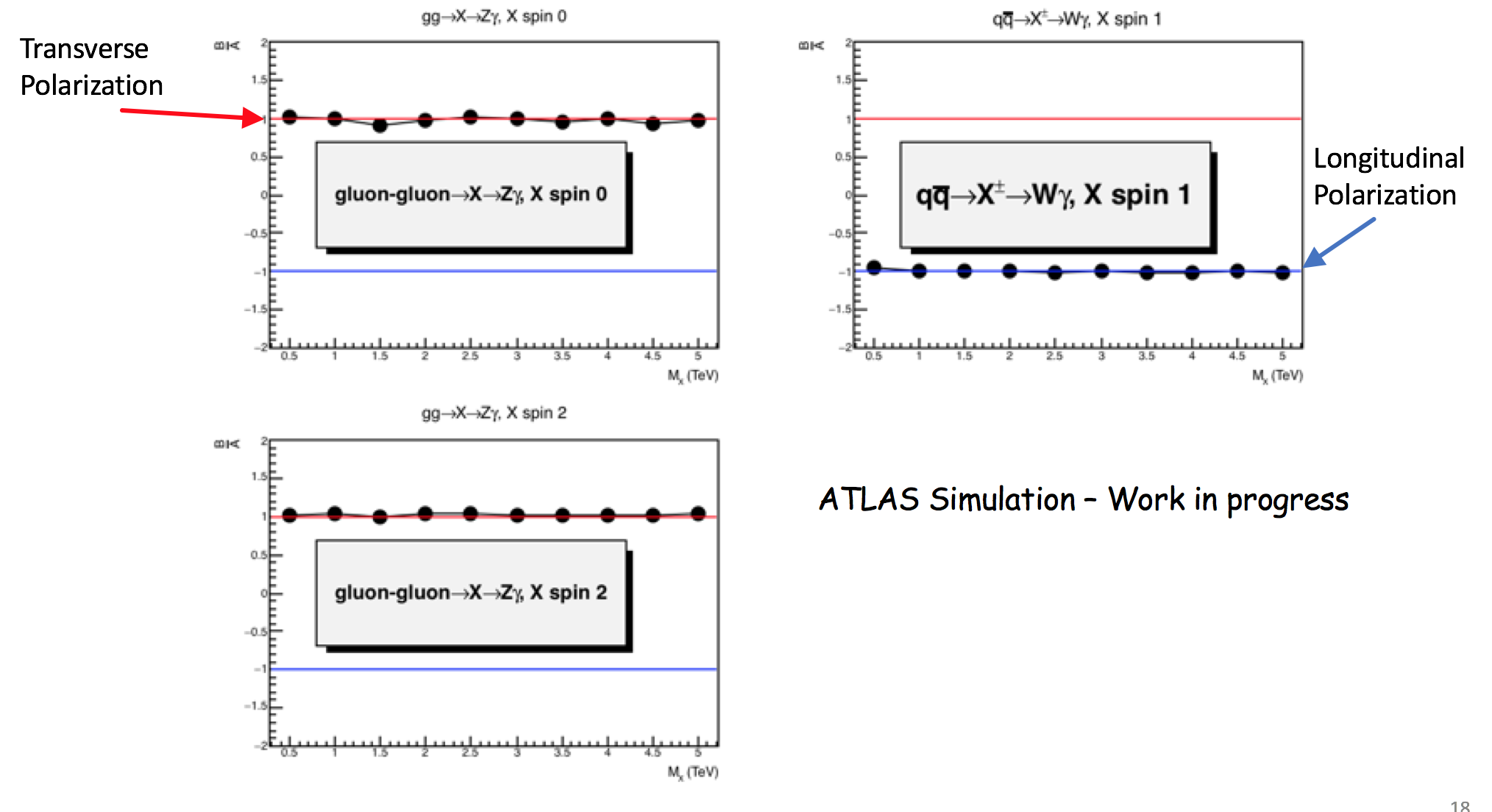, scale=0.45}
\caption[Caption for LOF]{$\frac{B}{A}$ polarization parameter for all $M_X$}
\label{ba}
\end{center}
\end{figure}

The identification of W/Z boosted jet is affected by $\Delta R(q\bar{q})$ distributions and hence the polarizations of the boson decay products. In order to quantify such decay, this paper defines a helicity frame and analyzes the $q\bar{q}$ decay angular polarizations in the frame.\\
Figure \ref{helicity} illustrates the definition of the helicity frame. The helicity frame is the W/Z boson rest frame with its Z axis defined to be along the same direction as the W/Z boson decay from the rest frame of massive boson X. We can reach the helicity frame by performing two consecutive sets of rotation and boost from the proton-proton rest frame. In the helicity frame, the $q\bar{q}$ decay products are back to back due to conservation of momentum. This paper studies the $\cos\theta$ distribution to quantify the polarizations.
Theoretically, the $\cos\theta$ distributions are different for the different W/Z production channels. Figure \ref{polarization_interpretation} shows the expected functional forms of the decay angular distributions from Standard Model electroweak decay of the W boson.\\
The $\cos\theta_{\bar{q}}$ distributions are fitted to the functional forms of $A+B\cos^2\theta=A(1+\frac{B}{A}\cos^2\theta)$. The fitting parameter $\frac{B}{A}$ is hence used to quantify the polarization. $\frac{B}{A}=1$ indicates a purely transverse polarization whereas $\frac{B}{A}=-1$ indicates a purely longitudinal polarization. Figure \ref{polarization_parameter} illustrates the distributions for purely longitudinal and purely transverse polarizations. Figure \ref{polarization_example} shows an example of $\cos\theta$ distributions at $M_X$=5TeV. It clearly demonstrates that the spin 0 and spin 2 productions are much more transversely polarized, while the spin 1 production is much more longitudinally polarized. The same fitting process is repeated for all $M_X$ slices and the $\frac{B}{A}$ parameters as a function of $M_X$ are plotted in Figure \ref{ba}. Spin 0 and spin 2 productions are almost purely transversely polarized while the spin 1 production is almost purely longitudinally polarized.

We use fat jet selection to identify W/Z boosted jet. This is using a cone radius of $\Delta R_{q\bar{q}}<1$, which is defined to be
\begin{equation}
\Delta R_{q\bar{q}}=\sqrt{\Delta\eta^2_{q\bar{q}}+\Delta\varphi^2_{q\bar{q}}}
\end{equation}

\begin{figure}[H]
\begin{center}
\epsfig{file=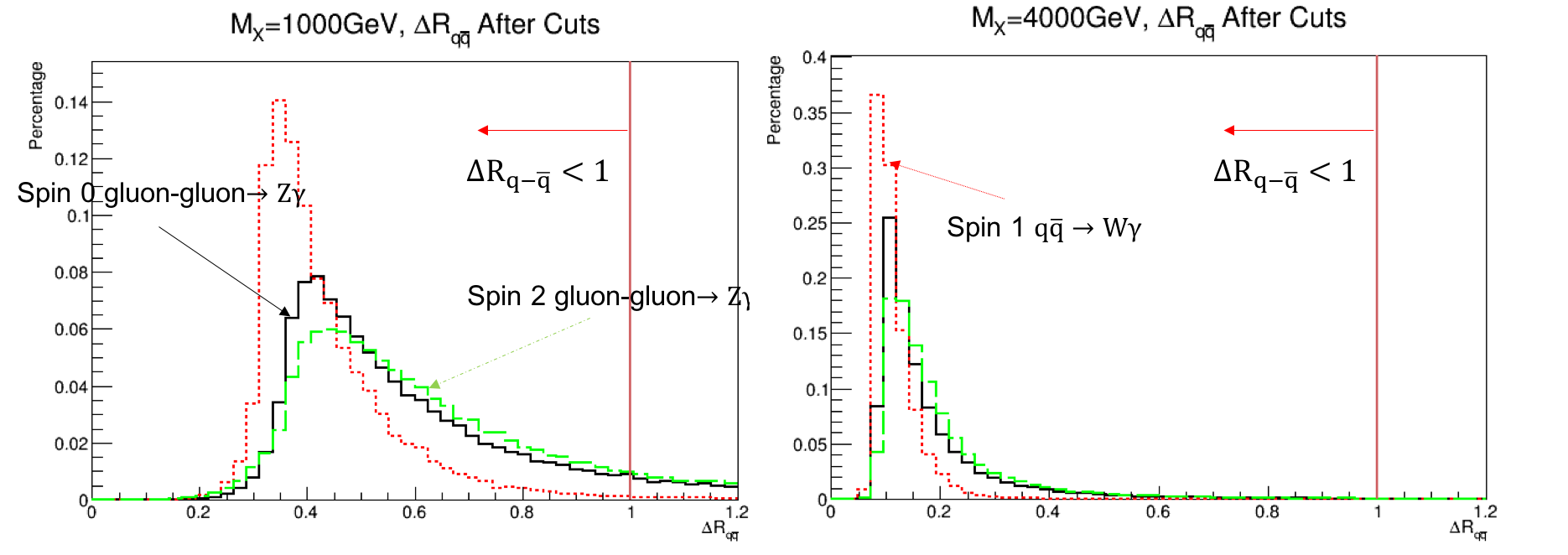, scale=0.45}
\caption[Caption for LOF]{$\Delta R_{q\bar{q}}$ distributions for $M_X=1TeV,4TeV$}
\label{delta_r}
\end{center}
\end{figure}

Figure \ref{delta_r} shows the $\Delta R_{q\bar{q}}$ distributions for two $M_X$ at 1TeV and 4TeV. We obtain good containment on the parton level. However, real data has extra inefficiencies due to presence of jets.

\section{Conclusion}
The ATLAS collaboration is carrying out a broad search for high mass bosons decaying to $Z\gamma$  and $W\gamma$ final states. Preliminary results using 36.1$fb^{-1}$ of data have been analyzed and place cross section limits on spin 0 production of $X\rightarrow Z\gamma$ with Z decays to $e^+$, $e^-$ and $\mu^+$, $\mu^-$ pairs. The limits for $\sigma_X\times BR(X\rightarrow Z\gamma)$ range between 88 fb for $M_X$ = 250 GeV and 2.8 fb for $M_X$ = 2.4 TeV.\\
The search sensitivity will be improved by including highly boosted W/Z bosons with hadronic decays. Preliminary studies have been done to determine selection cuts for spin 0 and spin 2 $X\rightarrow Z\gamma$  and spin 1 $X\rightarrow W\gamma$ decays. These show the potential for extending the search sensitivity for $X\rightarrow V\gamma$  decays for higher X masses.

\end{document}

%% file: econfmacros.tex
\def\beq{\begin{equation}}
\def\eeq#1{\label{#1}\end{equation}}
\def\eeqn{\end{equation}}

\def\beqa{\begin{eqnarray}}
\def\eeqa#1{\label{#1}\end{eqnarray}}
\def\eeqan{\end{eqnarray}}

\let\bar=\overbar

\def\Dslash{\not{\hbox{\kern-4pt $D$}}}
\def\dslash{\not{\hbox{\kern-2pt $\del$}}}

\def\msb{{\bar{\ssstyle M \kern -1pt S}}}

